# *Modeling the spreading of large-scale wildland fires*

## Mohamed Drissi [a]

a. University of Corsica, UMR CNRS 6134 SPE, Forest Fire Research Team, Campus Grimaldi, BP 52, 20250 Corte, France

**Abstract:**

The objective of the present study is twofold. First, the last developments and validation results of a semi-physical model designed to simulate fire patterns in heterogeneous landscapes are presented. Second, a sensitivity analysis was conducted to identify the most influential input model parameters controlling fire propagation. The model combines the features of a network model with those of a quasi-physical model of the interaction between burning and non-burning cells, which strongly depends on local conditions of wind, topography, and vegetation. Radiation and convection from the flaming zone, and radiative heat loss to the ambient are considered in the preheating process of unburned cells. The model is applied to an Australian grassland fire experiment as well as to a real fire that took place in Corsica in 2009. Predictions compare favorably to experiments in terms of rate of spread, area and shape of the burn. Finally, the sensitivity of the model outcomes (here the rate of spread) to six input parameters is studied using a two-level full factorial design.

**Nomenclature**

| | |
|---|---|
| $V$ | Control volume |
| $D$ | Cell diameter |
| $H$ | Cell height, Height |
| $WFF$ | Wet fine fuel |
| $DFF$ | Dry Fine fuel |
| $T$ | Temperature |
| $Q$ | Rate of heat release within the flame |
| $n$ | Number of sites |
| $t$ | Time |
| $i$ | Index of a burning cell |
| $j$ | Index of a non-burning cell |
| FMC | Fuel moisture content |
| FPPC | Fuel pyrolysis products content |
| $c_p$ | Specific heat capacity |
| $L_{vap}$ | Specific enthalpy change |
| a | Absorptivity |
| p | Elementary power carried by a quantum |
| $n"$ | Number of quanta emitted by each square meter of the flame |
| $P"$ | Emissive power of the flame |
| SNB | Small narrow-banded |
| MCM | Monte Carlo Method |
| SFM | Solid flame model |
| B | Stefan Boltzmann constant |
| $\dot{Q}$ | Rate of heat release within the flame |
| $H_{f0}$ | Flame height without wind |
| $H_f$ | Flame height with wind |
| S | Surface |
| $\Delta h_c$ | Heat combustion of volatile gases |
| $t_c$ | The residence time of the flame |
| $S_b$ | The superior surface of the site |
| U | Wind speed |
| L | Length |
| d | Distance |
| h | Mean convection coefficient |
| k | Thermal conductivity |
| $\vec{n}$ | Normal to the terrain (ground) |
| Re | Reynolds number |
| Pr | Prandtl number |
| G | Gravitational acceleration |
| X | Sensibility Factor |
| M | Matrix of levels |
| Ros | Rate of spread of fire |
| m | Mass |
| $m"$ | Mass load |

**Greek symbols**

| | |
|---|---|
| $\delta$ | Penetration length of radiation (optical length) |
| $\delta_{eff}$ | Effective penetration length of radiation |
| $\alpha$ | Volume fraction or compactness |
| $\sigma$ | Surface area/volume ratio |
| $\rho$ | Fuel particle density |
| $\varepsilon$ | Emissivity |
| $\alpha_f$ | Tilt angle of the flame |
| $\chi_R$ | Radiant fraction of the heat release within the flame |
| $\nu$ | Cinematic viscosity |
| $\kappa$ | Total absorption coefficient of the flame. |
| $\beta$ | Average extinction coefficient |

**Subscripts**

| | |
|---|---|
| k | Solid phase |
| ign | Ignition |
| WFF | Wet fine fuel |
| FDF | Fine dry fuel |
| AGL | Above ground level |
| fb | Fuel bed |
| $\infty$ | Ambient |
| str | Stratum |
| char | Char |
| f | Flame |
| ij | Between (i) and (j) |
| c | Combustion |

# Introduction

As revealed by satellite maps (Caldarelli *et al.*, 2001), fire patterns may exhibit irregular and even fractal shapes that are due to the local heterogeneous conditions (weather, fuel, and topography) encountered by the fire as it propagates.). The fire spread is multi-physical phenomenon involving different scales. At the gigascopic scale, the fire front is seen as a line separating the burned fuel from the virgin fuel, moving on a relief. But the essence of fire spread occurs at the macroscopic level: all the heat transfer mechanisms, the thermal degradation and the combustion take place at this scale. Moreover, it is at this level that the different local heterogeneities such as meteorology, topography and fuel interact with the fire front. The models developed at the macro-scale are combustion models (Grishin, 1987; Mell et al., 2007) which are based, for the most part, on a two-phase description of the flow and have a great generality. Most of the basic phenomena that govern the emergence and spread of wildfire are generally considered at the macro-scale: gas transport, turbulence, thermal degradation of the material, turbulent combustion, radiation exchange between the gas and the solid medium. A detailed model (Larini et al.,1998; Linn et al.,1997) requires the numerical solution of the balance equations of mass, energy and momentum for each phase and the radiative transfer equation. A realistic simulation therefore requires solving a system of equations involving tens of nonlinear partial and strongly coupled equations. The resulting CPU time resolution limits applications to areas of small size, typically a few hundred meters in three directions. These detailed models must be considered as models of knowledge for understanding the behavior of the fire, to test alternative hypotheses or enrich the models, but are not suitable for simulating the spread of fire at the large scale (gigascopic). At this scale, we deal with propagation models. There were different classifications of wildland models (Karplus, 1977; Pastor, 2003; Sullivan, 2007). Following Sullivan's classification, the macroscopic model developed here is quasi-physical since it doesn't attempt to represent all chemistry involved in fire spread but attempts only to represent the main physical phenomena. For the propagation model, it's common to treat the fire perimeter when making simulation as a group of contiguous independent cells that can grow in number, which is described in literature as a raster implementation (Sullivan, 2007). Thus, the propagation model takes the form of an expansion algorithm. Unlike algorithms of expansion based on direct-contact or on only nearest neighbours such as Cellular automata or percolation (*e.g.* Albinet *et al.*, 1986; Stauffer and Aharony, 1991; Ball and Guertin, 1992; Karafyllidis and Thanailakis, 1997; Hargrove *et al.*, 2000; Berjak and Hearne, 2002) the expansion algorithm used in this study expands the perimeter through considering long range effects between cells by taking advantage of the stochastic method of Monte Carlo to simulate radiative fluxes. We should notice the tentative of (Sullivan and Knight, 2004) to introduce a non-near neighbor spread to reproduce the parabolic head fire shape observed in experimental grassland fires (Cheney et al.,1993) but the ROS was not investigated. Graham, M and Matthai, 2003 followed by (Zekri et al., 2005) have introduced the concept of small world network in wildland fire to take into account the long range effect but no physics were incorporated. Recently, Adou et al., 2010 used the Monte Carlo method but the quasi-physical macroscopic model doesn't incorporate convection effect and the Monte Carlo method involves a circular neighborhood for each burning cell which is numerically a limiting procedure. The propagation model is based on elliptic expansion (Anderson et al., 1982) but the correlations of ellipses parameters for propagation are not physically based.

The present model investigates fire propagation on a landscape without making the assumptions of one-dimensionality spread (Catchpole et al., 1989; Weber,1991 ) neither of quasi-steady propagation (Dupuy and Larini, 1999). The vegetation and the flame are represented in three dimensions. The macroscopic model has emphasized on the phenomena that occur at the macroscopic scale, namely the preheating of the receptive fuel layer by the flame radiation and convection of hot gases, but also its radiative cooling to the surroundings. This is based on the resolution of an unsteady energy balance equation in every fuel element including detailed heat transfer mechanisms. The rate of spread (ros), the perimeter shape, the thickness of the fire front, the burnt area can be evaluated at each time.

The paper is organized as follows. First are introduced the general concepts of the model and the physical approach to determine the parameters upon which it depends. Second, results of the model are compared with real fire patterns. Third, a sensitivity analysis is conducted to investigate the effects of changes in certain model parameters on the fire spread rate.

## 1. The propagation model

### 1.1. The network

In this model, vegetation is depicted as a network of combustible cells that can be distributed on the soil surface either randomly or regularly depending on the coverage and spatial arrangement of vegetation, leading to either amorphous or crystalline networks. Vegetation coverage usually refers to the fraction of soil which is covered by vegetation, and spatial arrangement to the manner in which vegetation is distributed on the soil surface.

#### a. Crystaline network

In the case where vegetation is uniform and homogeneous, the network is assumed to be a crystaline network. For mono-disperse landscapes, the denser crystalline network may be constructed from triangular network where the cells are placed at the vertices of equilateral triangles and are in contact with each other (Figure1(a)). In this situation, vegetation coverage is maximum $f_{max} = \pi/2\sqrt{3} \sim 0.91$. A lower dense crystalline network can be constructed also on an underlying square-based structure. (Figure 1(b)).

#### b. Amorphous network

In the most cases, the vegetation is sparse and randomly distributed on the landscape. It seems more convenient to represent it by an amorphous network. This network should be generated with a predefined filling ratio that approaches the vegetation coverage rate. The amorphous network may be generated by randomly sowing the unit-size cells broadcast. This technique works well for a vegetation coverage of less than about 0.56. For denser networks, an optimization method based on a genetic algorithm is adopted (Figure1c). (Goldberg, 1989; Gen and Cheng, 1997; Back *et al.*, 1997).

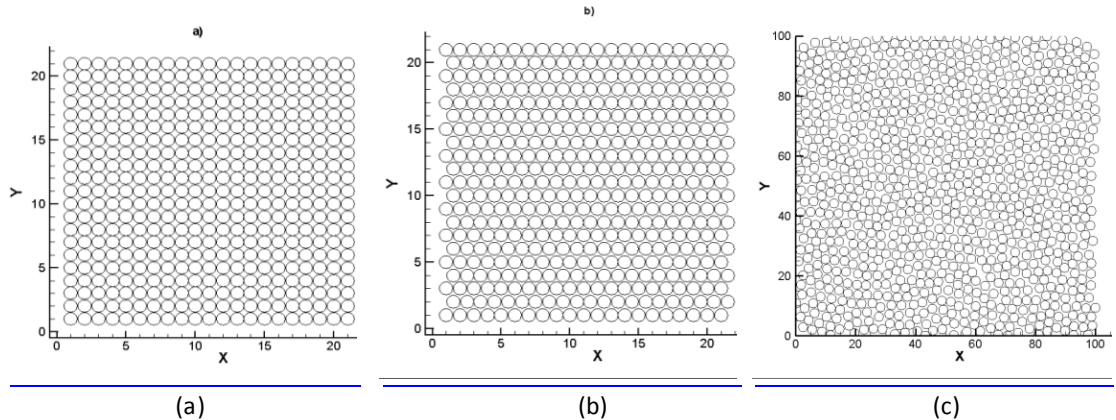

(a)           (b)           (c)

Figure 1: From left to right: examples of networks: a) Square based-structure network ,(b) Hexagonal based-structure network and (c) Amorphous network with a density of 70%..

## 1.2. The macroscopic model of combustion

The macroscopic model is based on an energy balance on a control volume about a virgin site including the radiant preheating from the front of flames as well as the convective preheating and the radiative cooling with the surroundings. It is an established fact that wildfire spread is dominated by the fine, thermally thin vegetative fuels (grass and foliage of shrubs and trees) (Rothermel, 1972), the spread of fire involves only the fine fuel elements while thicker elements burn more slowly at the back of the fire front. As suggested by (Luke and McArthur, 1977), fine fuels may be defined as organic material sufficiently small in size (*e.g.* fuel elements, typically less than 6 mm in thickness size (branches less than 6 mm in diameter). The physical problem is illustrated in Figure 2. Each combustible cell has a cylindrical shape with a height $H$ and a diameter $D$. In the present model, it is assumed that the elementary volume of the combustible cell involved in preheating is not the whole volume of the cell, but rather a top layer with a thickness $\delta$ and a volume $= \pi D^2 \delta \quad \delta (\delta \leq H)$. The thickness $\delta$, which cannot exceed $H$, corresponds to the mean free path of radiation through the vegetation medium. Beyond $\delta$, we assume that the medium no longer interacts with radiation. $\delta$It can be related to the surface-to-volume ratio of fine fuel elements, $\sigma_k$, and to the volume fraction of the solid phase, $\alpha_k$, as $\delta = 4/\sigma_k \alpha_k$ (De Mestre et al., 1989 ; Butler, 1993). As is commonly the case in fire models involving fine wildland fuels, the thermally thin assumption is adopted here, which means that there is no temperature difference in the control volume. This is assumed to apply if $\delta$ is small compared to flame length. The combustion model assumes that the thermal degradation until ignition occurs according to four stages: first the wet fine fuel elements (WFF) are heated until the boiling point of water, 373K, then the water evaporates, afterwards the dry fine fuel particles (DFF) are heated until the pyrolysis temperature and finally when the pyrolysis products content of the fuel ($FPPC$) reaches a some critical value, the vegetal ignites. This final stage is assumed to occur at the constant temperature of pyrolysis. Applying conservation of energy to the control volume $V_j$ of a receptive cell $j$ exposed to $N_{bc}$ burning cells yields

$$\sum_{i=1}^{N_{bc}}[q_{rad}^+(i) + q_{conv}^+(i)] = q_{rad}^-(j) + \begin{cases} \rho_{WFF} c_{p_{WFF}} \alpha_k \frac{dT(j)}{dt} & for\ T(j) < 373K \\ -\rho_{DFF} L_{vap} \alpha_k \frac{dFMC(j)}{dt} & for\ T(j) = 373K \\ \rho_{DFF} c_{p_{DFF}} \alpha_k \frac{dT(j)}{dt} & for\ 373\ K < T(j) < T_{pyr} \\ -\rho_{DFF} L_{pyr} \alpha_k \frac{dFPPC(j)}{dt} & for\ T(j) = T_{pyr} \end{cases} \quad (1)$$

Where $q_{rad}^+(i)$ and $q_{conv}^+(i)$ are respectively the energy per unit volume and per unit time received by cell $j$ from the burning cells ($i = 1\ to\ N_{bc}$) due to the energy-transfer mechanisms of preheating: flame radiation, wind-driven convection to the top surface of the receptive cell. For most nonzero ambient flow velocities, as is the case in the present study, other energy-transfer mechanisms of preheating, such as turbulent diffusion, solid- and gas-phase conduction, convective cooling and wind-driven convection within the cell, as well as the energy absorbed by pyrolysis prior to ignition, may be disregarded (Pagni and Peterson, 1973). The first term on the right-hand side of Eq.(1), $q_{rad}^-(j)$, represents the radiative loss from the fuel bed to the ambient surroundings. $L_{vap}$ is the specific enthalpy change of water to vapor at 373 K, $FMC(j)$ is the moisture content of $cell\ (j)$, on a dry basis ,i.e., the mass of water per the mass of dry fine fuel elements. Evaporation in fuel elements not at the boiling temperature is assumed to be negligible. $T(j)$ is the temperature of $cell\ (j)$, $\rho$ and $c_p$ are respectively the fuel particle density and specific heat of the solid phase. We assume that the volume of fine fuel elements remains constant during the drying process. $L_{pyr}$ is the latent heat of pyrolysis and FPPC is the content of pyrolysis products of fine dry fuel. The subscripts DFF and WFF refer to variables evaluated on a dry or wet basis. The initial FPPC is noted $FPPC_0$ and is defined as: $FPPC_0 = 1 - \nu_{char}$ where $\nu_{char}$ is the initial content of char of dry fine fuel elements. Each term of the energy balance equation is detailed below.

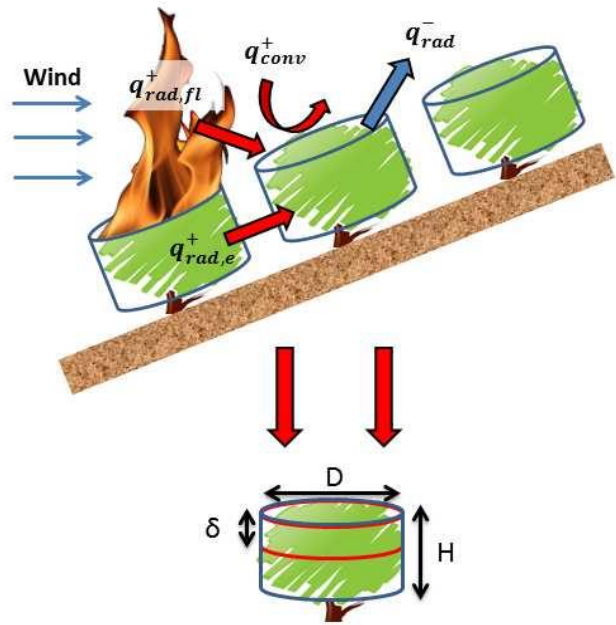

Figure 2: Flame spread schematic, with energy-transfer mechanisms indicated, and control volume of the cell involved in heat transfer.

a. **Radiation from the flame**

The amount of radiant energy received by the cell $j$ depends on flame emission, attenuation by the air layer between the flame and the cell, and absorption by the fuel medium. Overhead flame radiation is calculated by means of the Monte Carlo ray tracing method where the visible flame is regarded as a uniformly-radiating solid body with a cylindrical shape and with thermal radiation emitted from its surface. The limits of validity of this simple model of flame radiation, which is referred as to the "solid flame model", are discussed by Collin and Boulet (2013) in terms of the optical thickness of the flame. This radiation model is relatively simple, but it does require estimates of flame properties (*i.e.* height, length, angle, and base diameter) and emissive power . The Monte Carlo method is used to determine the number of quanta of energy launched from the burning cell (i) that finally reach the control volume of receptor cell (j). Each quantum carries an elementary power $p_i = P_i''/n''$ (in W), where $n''$ is the number of quanta emitted per unit flame area and $P_i''$ is the emissive power of the flame attached to the burning cell site (i) (see Figure The method is more detailed in (Adou et al., 2010). The radiative gain to cell j can thus be written as:

$$q_{rad}^+(j) = a_{fb}\, p_i N_{ij}/V_j$$

where $V_j$ is the control volume of cell j and $a_{fb}$ is the fuel medium absorptivity.

The contribution of burning cell *i* can be modified if there is another burning cell between cells i and j. This screening effect can easily be taken into account by the Monte Carlo method considering that a quantum emitted by cell *i* and crossing the solid flame attached to cell *k* is lost and does not contribute to increasing the energy of cell *j*. Moreover, radiation from the flame may be attenuated by the atmospheric layer between cells i and j. The probability of a quantum launched from a burning cell *i* in the direction *ij* being scattered or absorbed is determined from the Beer-Lambert law, using the transmittance of the atmospheric layer $\tau_{ij}$. This coefficient depends on the distance between the flame and the receptive cell, the relative humidity (RH) of ambient air, and the source temperature. It is calculated using the Statistical Narrow Band model of Malkmus (1967). For each emitted quantum a random number $R_s$ is generated to determine whether the quantum is absorbed or scattered ($R_s > \tau_{ij}$) by the atmospheric layer.

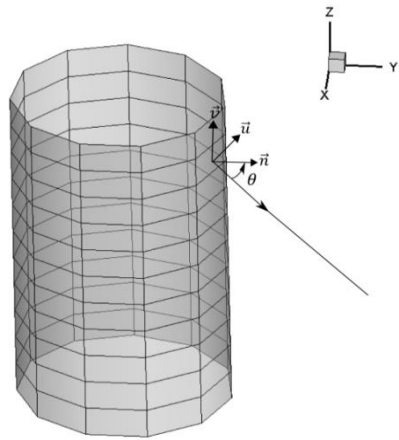

Figure 3: A quantum of energy emitted from the flame surface assumed to a cylinder in the context of solid flame model.

### b. Radiative losses

Unburned fuel elements within a cell (j) lose heat to the ambient by radiative heat loss at the top surface of the fuel bed: $q_{rad}^-(j) = \varepsilon_{fb}\sigma\left(T_j^4 - T_\infty^4\right)/\delta$ where $\varepsilon_{fb}$ is the emissivity of fuel bed, $\sigma$ is the Stefan Boltzmann constant ($\sigma = 5.67 \times 10^{-8}\,W/m^2/K^4$) and $T_\infty$ is the ambient temperature.

### c. Convection

As underlined by Beer (1990), the flame is not an impermeable barrier and, as a result of three-dimensional effects, the wind can penetrate the flame region.. Therefore, an unburned cell (j) located in the wake of a burning cell (i) can be heated by a convective flux in presence of wind. This flux depends on the distance between these two sites, $d_{ij}$, and decreases exponentially with a characteristic length of approximately three times the flame length (Pagni and Peterson, 1973)

$$q_{conv}^+(i) = \frac{h}{\delta}\left(T_f - T_j\right)e^{-0.3\,d_{ij}/L_f}$$

where $L_f$ and $T_f$ are the length and the temperature of the flame, $h$ is an average heat transfer coefficient empirically evaluated for a turbulent flow on a flat plate with a length of $L_f$ (Incropera and de Witt, 1985)

$$h = 0.037\,k\,Re^{0.8}Pr^{1/3}/d_{ij}$$

The Reynolds number, defined as $Re = U\,d_{ij}/\nu$, is based on the tangential component of the local wind velocity, $U = \left|\vec{n}\wedge(\vec{U}\wedge\vec{n})\right|$, where $\vec{n}$ is the normal to the ground surface. $Pr$, $\nu$ and k are respectively the Prandtl number, the cinematic viscosity and the thermal conductivity of air at the mean temperature $(T_f + T_\infty)/2$.

### 1.3. Input parameters of the propagation model

#### a. Flame properties

In this Section it is briefly explained how input model parameters, including input data, are determined. For a detailed description we refer to (Adou *et al.*, 2010). Following Putnam (1965), the luminous flame height under the influence of wind is estimated from

$$H_f = H_{f0}(1 + 4\,\frac{U^2}{gH_{f0}})^{-0.5}$$

where $H_{f0}$ is the luminous flame height without wind, deduced from the rate of heat release by combustion $\dot{Q}$, and the burning cell diameter, $D$, is given by Heskestad'correlation (Heskestad, 1983):

$$H_{f0} = 0.0148\,\dot{Q}^{2/5} - 1.02\,D$$

Since flaming combustion corresponds to the chemical reaction of volatiles with air, the associated heat release rate, $\dot{Q}\,(in\,W)$, may be expressed in terms of mass loss rate, $\dot{m}''_{DFF}\,(in\,kg.m^{-2}.s^{-1})$, the heat of combustion of volatiles $\Delta h_c\,(in\,J.kg^{-1})$, and the burning cell area, $S_{cell} = \pi D^2/4\,(in\,m^2)$, as

$$\dot{Q} = \dot{m}''_{DFF}\,\Delta h_c\,S_{cell}$$

As a first approximation, the pyrolysis rate may be related to the initial mass of DFF per unit area and flame residence time as $m''_{DFF} = \dot{m}''_{DFF}\,t_c$

Wind may affect flame geometry. Following Albini (1981), the tangent of the flame angle, defined as the angle between the flame axis and a vertical line, can be calculated as

$$\tan \alpha = 1.22\,U/\sqrt{gH_f}$$

However, due to the weak influence of wind on the flame length (Thomas and Pickard, 1961; Thomas, 1963; Nmira et al.,2010) , the flame length $L_f$ is assumed to be equal to the luminous flame height with no wind.

The emissive power per unit area of the flame can be expressed in terms of the flame surface area, $S_f = \pi D L_f$ , the radiant fraction of the chemical heat release rate lost by the flame, $\chi_R$, and the flame surface $S_f = \pi D L_f$ as $P''_f = \chi_R \dot{Q}/S_f$. So the temperature of the flame is deduced with the assumption that the flame is a grey emitting body with emissivity $\varepsilon_f$ so that: $T_f = \left(P''_f/\varepsilon_f\,\sigma\right)^{1/4}$. The emissivity of the flame can be expressed as a function of the effective total absorption coefficient $\bar{\kappa}$ of the flame and the length of the flame as $\varepsilon_f = 1 - e^{-\bar{\kappa}\,L_f}$. Following (Pard and Pagni, 1985) , $\bar{\kappa}$ is taken to be equal to $0.6\,m^{-1}$.

In applications where wind varies locally, wind speed and direction are calculated from average values using a numerical model (FLOWSTAR®) which takes into account the combined effects of the topography and surface roughness of the land site. This information with regard to roughness is based on the Corine Land Cover database (Silva *et al.*, 2007). The computational domain was large enough to ensure that boundary conditions did not affect the wind solution.

   b. **Fuel properties**

The fuel properties influence the way in which fire propagates. The table 1 shows the geometrical and thermo-physical properties of the fuel and of the fuel bed.

   c. **Meteorological conditions**

The environmental conditions are involved in the model through the ambient temperature, the relative humidity of air and mostly both the direction and the speed of the wind. The effect of the slope is taken into account through the MCM for radiation and through the tangential component (to the slope) of the wind for convection.

## 2. Model validation

The fire spread model was applied to two different fire scenarios: a grassland fire experiment in Australia (Cheney *et al.*, 1993, 1998) and an arson Mediterranean fire that occurred in Favone in Corsica in 2009 (Santoni *et al.*, 2011).

### 2.1. Australian grassland fire experiments

Results from grassland experiments conducted in Australia in 1986 (Cheney *et al.,* 1993, 1998) provide a useful set of data for model evaluation (see for example Mell *et al.*, 2005; Mell *et al.*, 2007; Sullivan, 2007). In the present work, model predictions are compared with observations from a grassland fire experiment, referred to as F19. In F19, the plot was 200m×200m and composed of *Themeda* grass (kangaroo grass). The fire was lit from 175m-long ignition line perpendicular to the prevailing wind on the upwind edge of the plot. This line fire was created with drip torches carried by two field workers walking for 56 s (87.5 m) in opposite directions from the center point to the ends the line fire. As observed by Cheney and co-workers, grass layers were continuous and homogeneous, which led us to use a triangular "crystalline" network, with a vegetation coverage of 0.91. The sizes of vegetation items were deduced from the correlation of Heskestad (1983) using the flame heights observed experimentally (Cheney *et al.*, 1993). The wind speeds at 2m above ground level at each corner of the experimental plot were measured at 5s intervals throughout the duration of the fire. Fire perimeters are plotted at times t=56s and t=86s. (Cheney *et al.*, 1993) in Figure 4. They are compared with those predicted by the model using the parameter values given in Table 1.

A good agreement is shown at times 56 s and 86 s. The average rate of spread of the head fire front from the first to the second contour is slightly overestimated, by less than 10 percent. Unfortunately, after change of 20° or more in the wind direction occurred (Cheney *et al.*, 1993), which renders the subsequent comparison unreliable. Another excellent indicator of fire behavior is the average width of the head fire front since this characteristic length is the product of the rate of spread and the flame residence time ($d = ros \times t_c$). The model gives approximately 9.4 m, whereas experiments gave about 10 m. It is worth noting that experimental mean values of the *ros* and head fire width leads to a mean flame residence time of 8.1 s, which is slightly higher than that deduced from the Anderson formula, 6.2 s. It is worth no ting also that the model based only on the description of radiative heat transfers from the flame (Adou et al., 2010) is not sufficient and can't predict correctly the fire spread, at least in the case of experiment F19. It was necessary to add the convective flux and the radiative losses to the ambient surroundings, which are non-negligible heat transfer contributions at the macroscopic scale of the fire front.

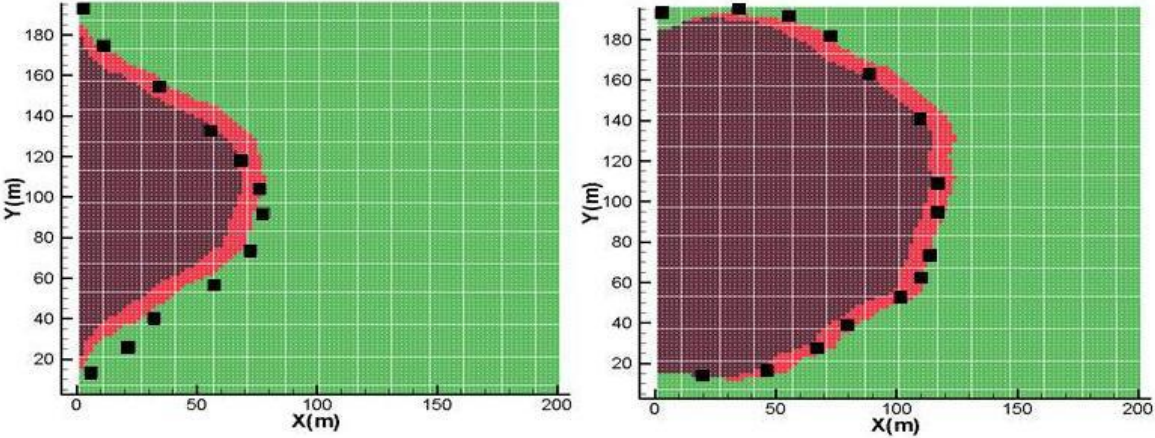

Figure 4: Fire contours predicted by the model (red contour) and measured (■) after 56s (at left) and 86s (at right) of propagation.

Table 1: Input model parameters for grassland fire experiment F19.

| **Input model parameters** | **Symbol (Unit)** | **Value** | **Source** |
|---|---|---|---|
| Ratio surface/ volume | $\sigma_k (m^{-1})$ | 12240 | (Cheney et al., 1993) |
| Char content/ content of gaseous pyrolysis products | $\nu_{char}/FPC_0$ | 0.20/0.80 | (Sussot [38],1982) |
| Specific heat | $c_{p,k}\ (J\ kg^{-1} K^{-1})$ | 1110+3.7*T | (Parker [39], 1989) |
| Stratum height | $H(m)$ | 0.51 | (Cheney et al., 1993) |
| Density of fuel particle | $\rho_k,\ kg\ m^{-3}$ | 512 | (Rothermel [35], 1972) |
| Dry load | $m''_{DFF}, kg\ m^{-2}$ | 0.313 | (Cheney et al., 1993) |
| Volume of solid phase fraction | $\alpha_k$ | 0.0012 | (Cheney et al., 1993) |
| Initial Moisture content | $FMC_0$ | 0.058 | (Cheney et al., 1993) |
| Pyrolysis temperature | $T_{pyr}, K$ | 500 | (Koo et al. [29], 2005) |
| Ignition temperature | $T_{ign}, K$ | 500 | (Koo et al., 2005) |
| Critical content of pyrolysis products | $FPC_{cr}$ | 0 | Present study |
| Radiated Fraction | $\chi_r$ | 0.35 | (Quintiere ,1997) |
| Heat of combustion | $\Delta h_c, (J.kg^{-1})$ | $15.6 \times 10^6$ | (Sussot 1982 ; Hough 1969 [40]) |
| Mean absorption coefficient of the flame. | $\kappa_f (m^{-1})$ | 0.4 | (Koo et al., 2005) |
| Residence time of the flame | $t_c(s)$ | 5 | (Koo et al., 2005) |
| Fuel bed absorptivity | $a$ | 0.9 | (Koo et al., 2005) |
| Flame height | $H_f(m)$ | 2.04 | (Cheney et al., 1993) |
| Wind speed (at 2m AGL) | $U(m.s^{-1})$ | 4.83 | (Cheney et al., 1993) |
| Relative humidity of the air | $RH(\%)$ | 20 | (Cheney et al., 1993) |
| Cell Diameter | $D(m)$ | 2.54 | Deduced from Heskestad's correlation |
| Ambient Temperature | $T_\infty (K)$ | 307 | (Cheney et al., 1993) |

## 2.2. Corsican Fire

This fire has been extensively studied and documented by Santoni *et al.* (2011). Geographical data include a digital elevation model and vegetation map at a resolution of 25 m. This case is of interest because the fire propagated first upslope, between points A and B (Figure 5), and then downslope. The dominant vegetation was composed of a mixture of foliage of live strawberry shrubs (*Arbutus Unedo*) and leaf litter. The weather conditions at the time of the fire were: an average wind speed of 6m.s$^{-1}$ at 10m above ground level, an average direction of 270° (west), and a dry bulb temperature of 303 K. Since the landscape was almost continuously covered with homogeneous vegetation, a triangular "crystalline" network is used, with a maximum vegetation coverage of 0.91. Comparison between predicted and real fire contours shows that the rate of spread of the head fire front is well predicted by the model (Figure 5).

After 29 min of upslope fire propagation (point B in Figure 5), the average rate of spread of the head fire front is slightly underestimated, 14.0 vs. 16.8 m.min$^{-1}$. The same trend is observed in the downslope part of the terrain, between points B and C, with 10.8 vs. 13.3 m.min$^{-1}$. On the other hand, the post-fire burned area predicted by the model is overestimated, 34 vs. 29 ha. Lateral expansion discrepancies are observed which may be due to the fire crew intervention on the fire's flanks. This was not introduced into the model because of lack of information on the exact location and nature of the firefighting task force deployed. Possible errors in the estimation of some input parameters may also explain these discrepancies.

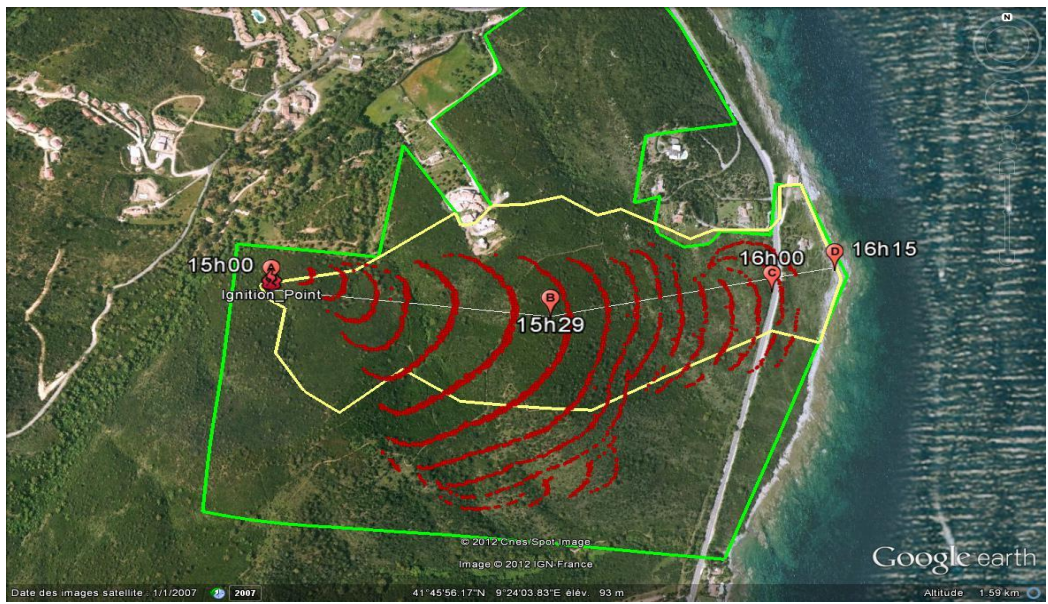

Figure 5: Predicted 5-min fire contours (red lines) and post-fire GPS recordings (yellow line). Measured times of the fire front passage at points B and C are indicated (Santoni et al.,2011).

# 3. Sensitivity analysis

The fire spread model presented above is based on the knowledge of the parameters given in Table 1. Its predictive capability depends on the accuracy with which these parameters are determined. Unfortunately, most of them are difficult to measure and/or exhibit a high degree of variability, which in turn makes the output uncertain. The aim of the sensitivity study is to answer the following questions: (1) which of the model parameters exert the greatest influence on model outputs, and thus should serve as guides to future research in the field; (2) which parameters are insignificant and can be eliminated from the final model or replaced by a nominal value (model reduction). Sensitivity analysis can be performed either comprehensively or just partially by considering selected parameters only, a few parameters that we judge *a priori* to be important. An excellent review of the different methods employed to conduct a sensitivity analysis is given by Hamby (1994). The current analysis is conducted on a flat terrain of 100m × 100m where the vegetation is represented by a triangular network of cells having each a diameter of 2 m, and a height of 2.5 m. We have used here as input parameters of the fuel, the thermo-physical properties and the geometrical properties of Kermes Oak (Drissi, 2013). The sensitivity analysis is based on a full factorial design, involving six input parameters or factors, each at two levels ($x_i = \pm 1$). This factorial design thus requires $2^6 = 64$ model runs. Table 2 presents the six factors considered and their range of variation. The range of parameter variations is somewhat arbitrary but it is broad enough to cover typical fuel and meteorological conditions. The range of variation for the ignition temperature $T_{ign}$ is smaller than for the other factors in order to avoid undesirable fire scenarios (no propagation or incomplete propagation), which render the sensitivity analysis meaningless. This was observed for $T_{ign} < 530K$ depending on the values of the other parameters. Here, the response of the rate of fire spread associated with the dynamics of the fire is considered: here computed by dividing the length of the domain by the time required for the head fire front to move from the ignition location to the opposite end of the domain.

Table 2: Parameters of the full factorial design and range of variations.

| Parameter | Reference level | Low level (-1) | High level (+1) |
|---|---|---|---|
| Dry load of fine fuel elements $m''_{DFF}$ ($kg.m^{-2}$). | 3.0 | 2.5 | 3.5 |
| The residence time of the flame $t_c(s)$ | 30 | 27 | 33 |
| Initial moisture content $FMC_0$ | 0.2 | 0.16 | 0.24 |
| The ignition temperature $T_{ign}(K)$ | 550 | 540 | 560 |
| Fraction of heat loss by radiation $\chi_r$ | 0.5 | 0.45 | 0.55 |
| Wind speed ($U$ ($m.s^{-1}$) | 5 | 4 | 6 |

The main effects of parameters on the burning area are shown in Figure 6. It can be observed that the wind speed has the greatest effect on the rate of fire speed, followed by the load of fine fuel elements, and to a lesser extent by the flame time residence, the radiative heat loss fraction, the initial moisture content and the ignition temperature.

The Pareto's chart confirms these observations (figure 7). The factors with amplitude that exceeds the red line are the more significant and strongly affect the rate of fire spread. One can observe that main effects are dominant, followed by some second-order interaction effects. The most influential interaction effect on the ros is due a two-factor interaction (AF in figure 7) between the load of fine fuel elements and the wind speed. Third-factors interactions do not significantly affect the ros.

Another advantage of the sensitivity study is that correlations may be established to estimate responses as functions of these parameters. The method consists in approximating the response of the system, here the head fire rate of spread, by the following expression

$$ros \sim ros_0 + \underbrace{\sum_{i=1}^{6} \beta_i X_i}_{\substack{main\,effects\\order\,1}} + \underbrace{\sum_{i=1}^{6}\sum_{j>i} \beta_{ij} X_i X_j}_{\substack{interactions\\order\,2}} + \underbrace{\sum_{i=1}^{6}\sum_{j>i}\sum_{k>j} \beta_{ijk} X_i X_j X_k}_{\substack{interactions\\order\,3}}$$
$$+ \underbrace{\sum_{i=1}^{6}\sum_{j>i}\sum_{k>j}\sum_{l>k} \beta_{ijkl} X_i X_j X_k X_l}_{\substack{interactions\\order\,4}} + \underbrace{\sum_{i=1}^{6}\sum_{j>i}\sum_{k>j}\sum_{l>k}\sum_{m>l} \beta_{ijklm} X_i X_j X_k X_l X_m}_{\substack{interactions\\order\,5}}$$
$$+ \underbrace{\beta_{123456} X_1 X_2 X_3 X_4 X_5 X_6}_{\substack{interactions\\order\,6}} \qquad (2)$$

where $ros_0$ is the rate of spread predicted by the model using reference values (Table 2). The mathematical system can be written $\vec{ros} = X.\vec{\beta}$ where $\vec{ros}$ is the response vector, $X$ the 64×64 level matrix, and $\vec{\beta}$ the vector of coefficients. $\vec{\beta}$ is calculated from $\vec{\beta} = X^{-1}\vec{ros}$ using a LU decomposition method. The main effects $\beta_i$ is the overall effect of the parameter i on the ros, $\beta_{ij}$ is the interaction effect between the factors i and j, etc.

Ignoring second-order and higher-order terms in Eq (2) leads to the first-order approximation of the rate of spread,

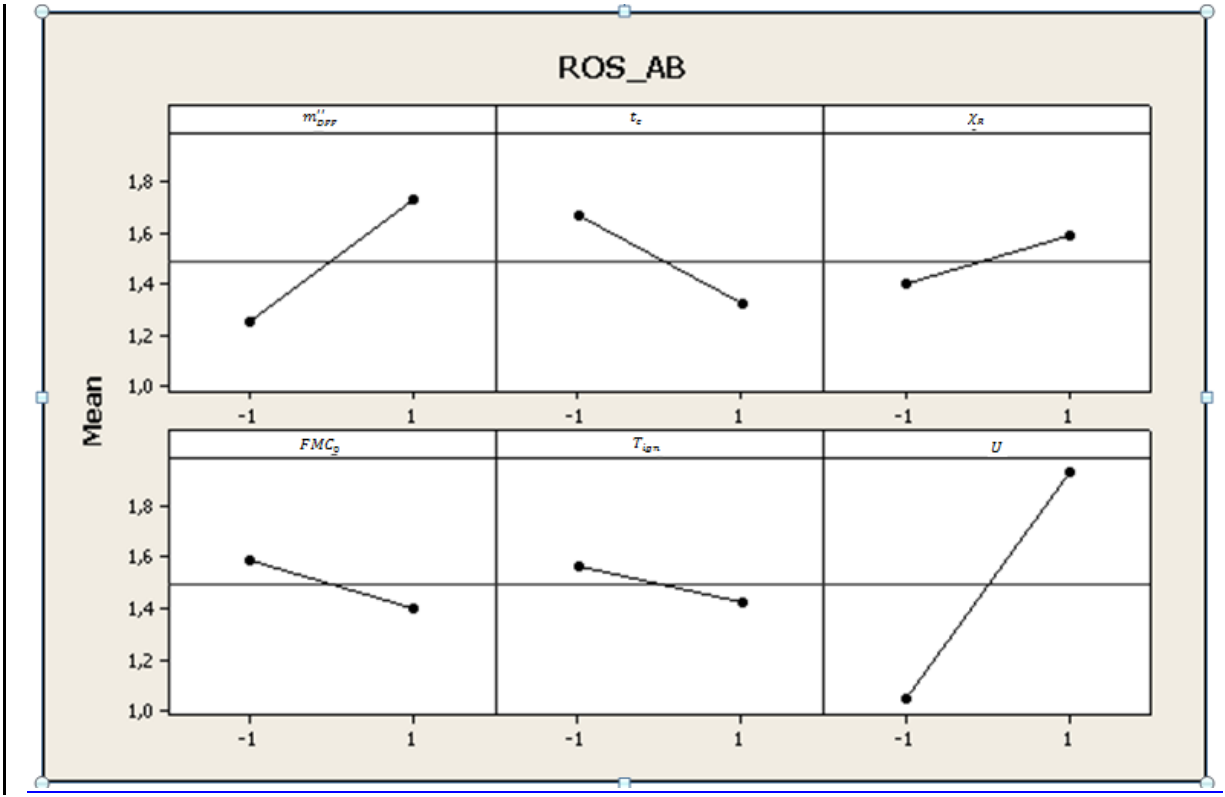

Figure 6: Main effects of parameters on the mean of fire spread rate.

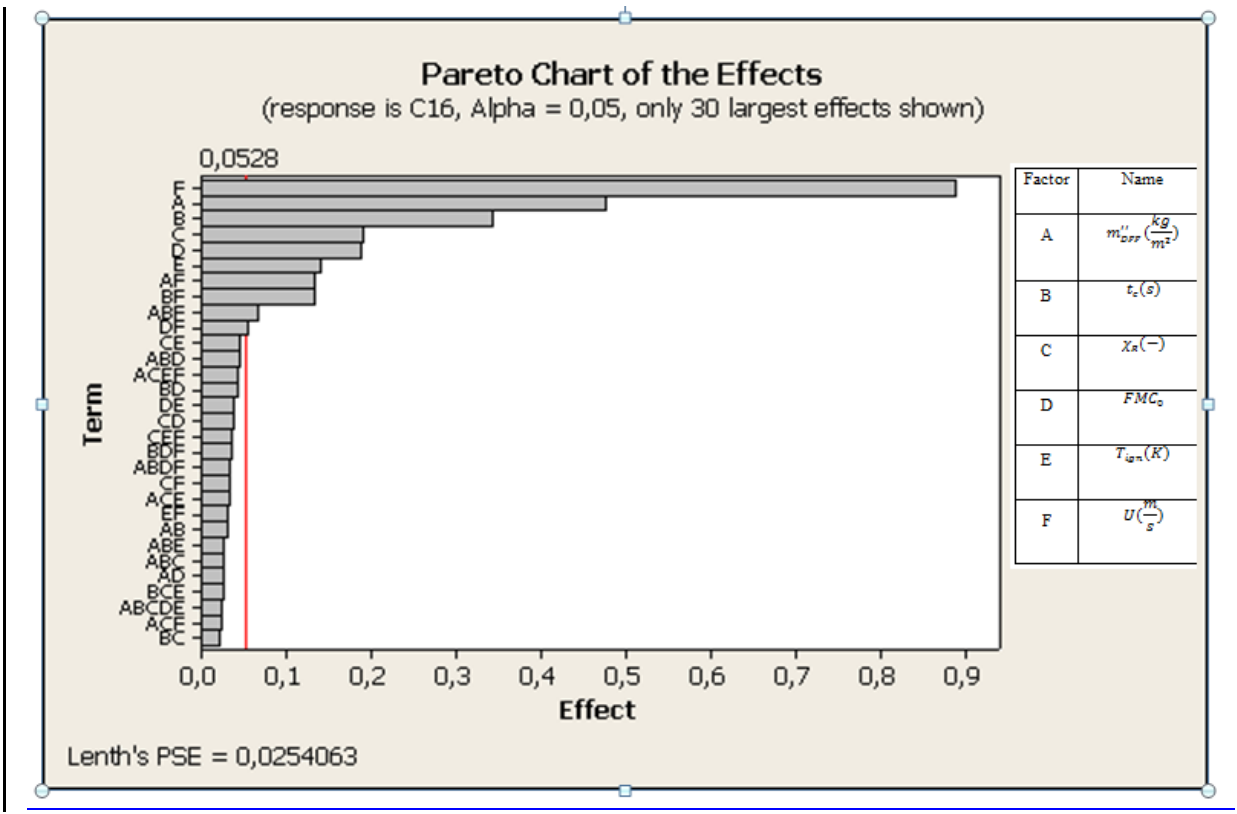

Figure 7: Pareto's chart of the effects of parameters on the mean of fire spread rate.

$$ros_1 \sim ros_0 + \underbrace{\sum_{i=1}^{6} \beta_i X_i}_{main\,effects\,order\,1}$$

Or in terms of physical parameters,

$$ros_1 \sim 1.4907 + \frac{0.2388}{0.5}(m''_{DWF} - 3) - \frac{0.1716}{3}(t_c - 30) + \frac{0.0955}{0.05}(\chi_r - 0.5)$$
$$- \frac{0.0939}{0.04}(FMC_0 - 0.2) - \frac{0.0707}{10}(T_{ign} - 550)$$
$$+ \frac{0.444}{1}(U - 5) \quad (3)$$

Equations (2) and (3) may be also used to quantify the change in the ros as a function of any of the model parameters. This fire spread rate can be compared with simple correlations from the literature (Morvan et al., 2009; Overhalt et al., 2014). The relative errors made when calculating the ros using first and second-order approximations are plotted for the 64 runs of the full factorial design (Figure 8).

It is found that the relative error for the first-order approximation varies approximately between 10% and 30% , whereas it drops to about 10% using the second-order approximation. This means that Eq (3) may be used as a good approximation to evaluate the ros for any fuel and meteorological conditions that fall within the range of variations given in Table 2.

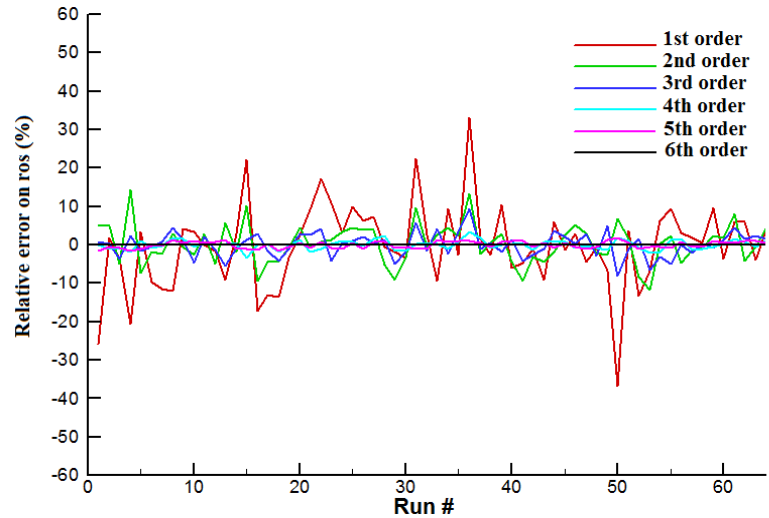

Figure 8: Relative errors of first and second-order approximations of the ros.

## 4. Conclusion

The latest improvements of a semi-physical network model of wildfire spread are presented. Vegetation is depicted as a "crystalline" or an amorphous network of combustible cells depending on the fuel map On a macroscopic scale, the preheating energy transfer mechanisms considered are: flame and ember radiation, wind-driven convection and radiative heat loss to the ambient. The effects of wind, topography, and vegetation are included in the model. Model results compare favorably with data from a real fire that occurred in Corsica, and Australian grassland. In the case of grassland fires, a good concordance is found for both the head fire rate of spread and the width of the fire front. It appears that radiation is the dominant preheating mechanism, with a significant contribution of convection. Moreover, the network model provides super-real-time simulations, five to ten times faster than real time, and thus can be used as the basis of a fire-fighting decision support system. A sensitivity study is performed using a full factorial plan of experiment showing how sensitive the rate of fire are to variations in certain model parameters. This could help to identify the parameters we should focus on in the future. Analysis results are used to derive correlations between the responses and model parameters. Such correlations may be used to give rapid estimates of the rate of spread of a fire with fuel and meteorological conditions which depart from reference values.

**Acknowledgments**

I would like to thank the SPE team in Corsica and especially Professor P.A. Santoni for making available the data concerning the Favone Fire. I would thank also Professor Noureddine Zekri for fruitful discussions about the topic. I would thank the "Ministère de l'Education Nationale , de l'Enseignement Supérieur et de La Recherche , France" for the funding of this research.